\documentclass{article}
\usepackage{spconf,amsmath,graphicx}
\usepackage{amssymb}
\usepackage{xcolor}
\usepackage{hyperref}

\title{A MULTI-CHANNEL TEMPORAL ATTENTION CONVOLUTIONAL NEURAL NETWORK MODEL FOR ENVIRONMENTAL SOUND CLASSIFICATION}
\name{You Wang, Chuyao Feng, David V. Anderson}
\address{Georgia Institute of Technology, School of Electrical and Computer Engineering, Atlanta, Georgia, USA}
\begin{document}
%\ninept
%
\maketitle
\begin{abstract}
Recently, many attention-based deep neural networks have emerged and achieved state-of-the-art performance in environmental sound classification. The essence of attention mechanism is assigning contribution weights on different parts of features, namely channels, spectral or spatial contents, and temporal frames. In this paper, we propose an effective convolutional neural network structure with a multi-channel temporal attention (MCTA) block, which applies a temporal attention mechanism within each channel of the embedded features to extract channel-wise relevant temporal information. This multi-channel temporal attention structure will result in a distinct attention vector for each channel, which enables the network to fully exploit the relevant temporal information in different channels. The datasets used to test our model include ESC-50 and its subset ESC-10, along with development sets of DCASE 2018 and 2019. In our experiments, MCTA performed better than the single-channel temporal attention model and the non-attention model with the same number of parameters. Furthermore, we compared our model with some successful attention-based models and obtained competitive results with a relatively lighter network. 

\end{abstract}

\section{Introduction}
\label{sec:intro}

 Environmental sound classification (ESC) is an important area of sound event detection and classification with applications in many areas including  audio surveillance systems \cite{foggia2015audio}, hearing aids \cite{alexandre2007feature}, smart rooms \cite{vacher2007sound}, and smart cars, etc. %Environmental sounds are a very diverse group of everyday audio events that can neither be described as speech nor as music \cite{piczak2015esc}. In recent years, a number of datasets containing various classes of environmental sounds and acoustic scenes have become available, and among them the most widely used ones includd ESC-50 \cite{piczak2015esc}, UrbanSound8k \cite{Salamon:UrbanSound:ACMMM:14}, DCASE datasets \cite{mesaros2018detection, mesaros2017dcase}, and Google Audioset \cite{gemmeke2017audio}. 
Traditional machine learning algorithms for audio pattern recognition include K-nearest neighbors, support vector machines, and Gaussian mixture models, etc. But with the support of more labelled datasets, neural networks based methods including convolutional neural networks \cite{piczak2015environmental, salamon2017deep}, recurrent neural networks \cite{ren2017deep, phan2017audio}, and their combination \cite{xu2018large, zhang2019attention}, have achieved superior performance over the traditional approaches. 

In recent years, an attention mechanism was introduced to emphasize certain parts of the feature set to further improve the performance. The concept of attention was first introduced in \cite{bahdanau2014neural} for machine translation. Since then, attention-based neural networks have been widely used in computer vision \cite{woo2018cbam}, natural language processing \cite{wang2016attention} and speech recognition \cite{chorowski2015attention}. 
In audio classification, attention mechanisms are used to assign importance to channels, spectral or spatial content, and/or temporal frames. Because of the unique temporal structure of audio signals, temporal attention has been very widely used for audio classification. For general audio scene classification, this is particularly appropriate since salient characteristics of an acoustic scene may be only briefly present or may reside only in certain features.   One goal of the many proposed attention mechanisms is to discover the best approach for identifying the salient regions or features.  
Yu \textit{et al.} \cite{yu2018multi} proposed a multi-level attention model for weakly labelled audio classification % it looks like we are missing a word here--what is weakly referring to?
that applies temporal attention to single-channel embedded feature maps. Li \textit{et al.} \cite{li2019multi} proposed a multi-stream network with temporal attention in which the structure is composed of three streams, each containing a single temporal attention vector. %Zhao \textit{et al.} introduced a spatial attention pooling model for DCASE 2018 \cite{ren2018attention}, and then further combined the idea with Atrous CNN \cite{ren2019attention}. 
Zhang \textit{et al.} \cite{zhang2019attention} integrated temporal attention into its CRNN architecture and the same authors \cite{zhang2019learning} proposed a model that combines channel attention and temporal attention together.    

However, the above audio classification models either do not embed input feature maps into multiple channels or apply only one single temporal attention vector on feature maps of all channels. Therefore, these models fail to take advantage of the fact that feature maps in different channels actually have different temporal structures. To overcome this limitation, we propose a CNN model with a multi-channel temporal attention (MCTA) block that extracts different temporal attention vectors for different channels to more fully exploit channel-wise temporal information.

% \begin{figure}[ht]
% \includegraphics[scale=0.31]{example.png}
% \caption{\small 3-channel input of a sneezing sound. Row number 1-3 represents mel-spectrogram, deltas, and delta-deltas respectively. The highlighted red rectangles are the temporally relevant parts.}
% \label{fig:example}
% \end{figure}

\section{Multi-Channel Temporal Attention (MCTA) CNN Model}
\label{sec:proposed}

Typically, a CNN will apply convolution on the input with multiple filters to extract more high-level channel-wise features. These channels contain information from different aspects of the input, all of which will contribute to the final classification. Our proposed MCTA model provides each channel with a unique attention vector, better exploiting the different structures of channel-wise feature maps and different information associated with them.

\begin{figure*}[ht]
\centering
\includegraphics[scale=0.12]{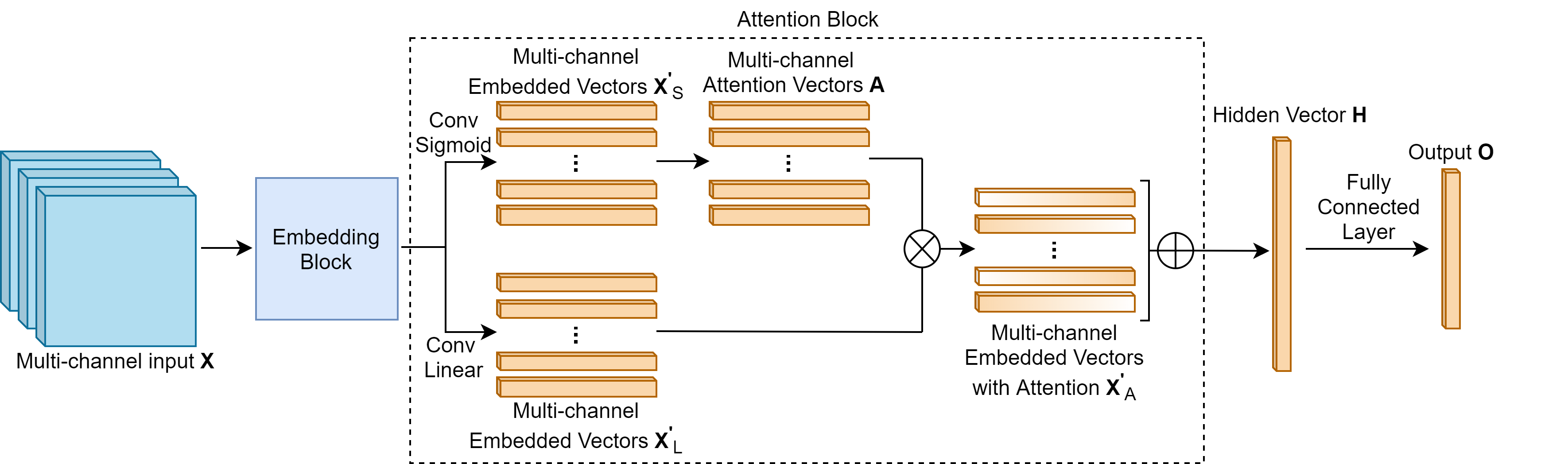}
\caption{\small The block diagram of our proposed multi-channel temporal attention model. The $\otimes$ sign stands for Hadamard product and the $\oplus$ sign represents summing over the time dimension and squeezing the frequency dimension.}
\label{fig:architecture}
\end{figure*}

\subsection{Model Architecture}
The architecture of our proposed model is shown in \textbf{Fig. \ref{fig:architecture}}. The model takes multi-channel feature maps $\mathbf{X}\in\mathbb{R}^{C\times{T}\times{F}}$ as input where $C$ is the number of channels, $T$ is number of time frames and $F$ is number of mel-frequency bins, and passes them through an embedding block that consists of several convolutional and max-pooling layers. For example, the feature map extracted from ESC-50 dataset has a size of $3\times{431}\times{128}$. % do you mean that each sample from the ESC-50 dataset has 431 frames?
The embedding block will embed the input to a hidden number of channels and aggregate it along the frequency dimension, yielding a set of embedded feature vectors $\mathbf{X}^\prime\in\mathbb{R}^{C^\prime\times{T^\prime}\times{1}}$, where $C^\prime=512$ and $T^\prime=52$ in the ESC-50 case. These embedded vectors then enter the attention block where the network starts to bifurcate to generate the further embedded vectors $\mathbf{X}_{L}^\prime$ and the attention vectors $\mathbf{A}$ respectively, and then the attended vectors $\mathbf{X}_{A}^\prime$ are obtained by performing element-wise multiplication. Finally the time and frequency dimensions of $\mathbf{X}_{A}^\prime$ are both squeezed to form a hidden vector $\mathbf{H}\in\mathbb{R}^{C^\prime\times{1}}$, which is then passed through a fully connected layer for the final classification.

\subsection{The Embedding Block}
\begin{figure}
\centering
\includegraphics[scale=0.13]{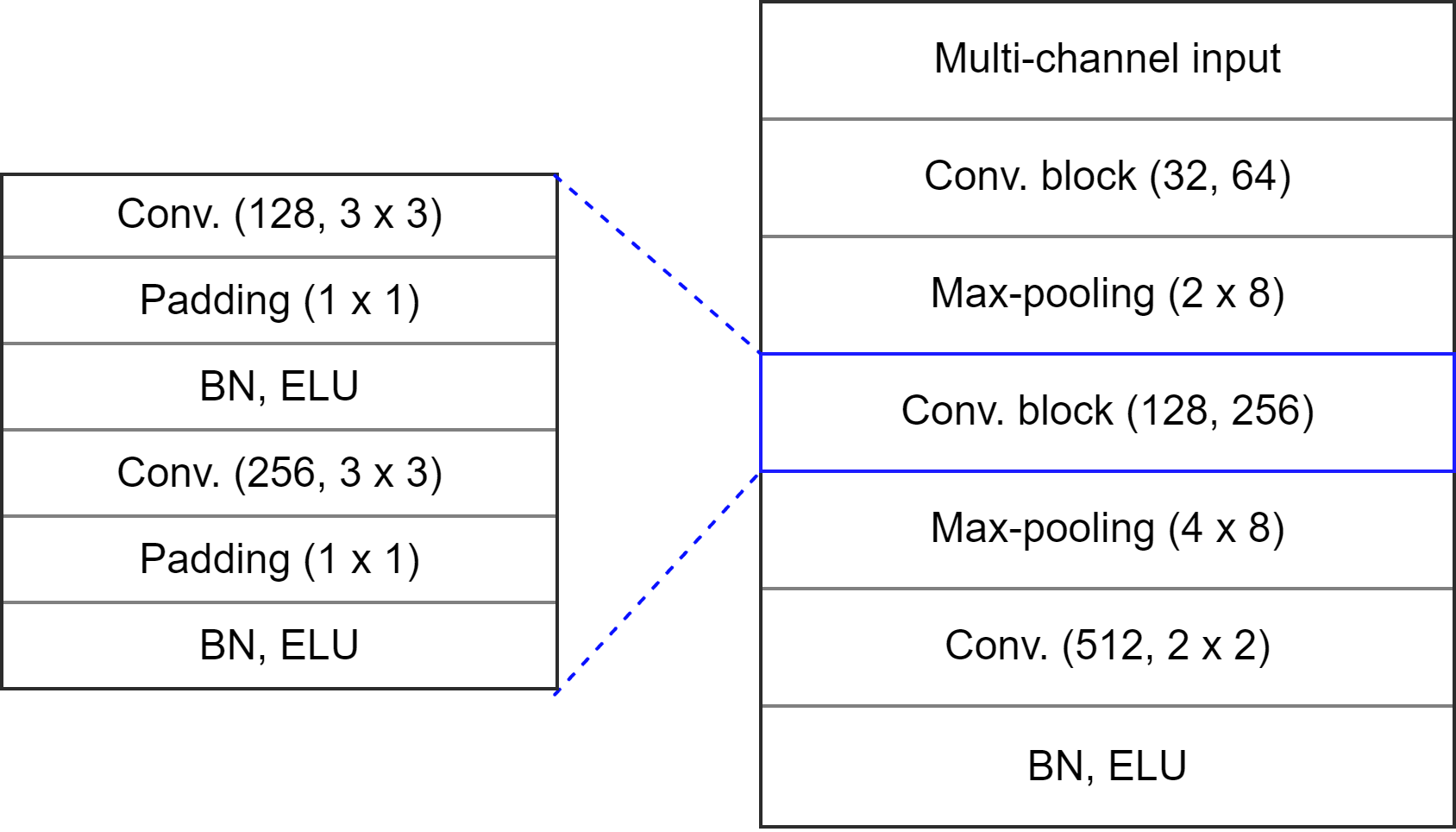}
\caption{\small Details of the embedding block.} 
\label{fig:embedding}
\end{figure}

\textbf{Fig. \ref{fig:embedding}} presents the details of the embedding block. Because of the size of the datasets used in our work, a relatively shallow CNN structure is used, having a total of 5 convolutional layers and 2 max-pooling layers. Inspired by the work of Kumar \cite{kumar2018knowledge}, our embedding block is designed to eventually aggregate the 128 mel bins to 1 and reduce the time dimension by max pooling to extract useful segment-level temporal information.  After each convolutional layer, batch normalization (BN in the figure) is performed and ELU (Exponential Linear Unit) is used as activation. The numbers in the brackets after ``Conv. block''  indicate the numbers of filters of the two convolutional layers in them. The numbers after ``Max-pooling'' and ``Padding'' stand for kernel sizes. The numbers after ``Conv.'' are the numbers of filters and the kernel size. %Therefore, our model is more suitable for smaller datasets and datasets without pre-extracted features and we believe that ``shallower'' embeddings can take more advantage of the attention mechanism before they become too deep.
% For the convolutional layers in all the three blocks, kernal size is $3\times3$ with stride and padding fixed to 1, and the 2d max-pooling is done with both kernal size and stride as $2\times4$. Block B4 that marks the end of the embedding module only consists of one 2d-convolutional layer also with batch-normalization and stride as 1, but the kernal size is $2\times2$ with zero padding. RELU activation is applied on all the convolutional layers. Following the embedding module is the attention module, where the network splits into two branches as block A1 and H1. Both A1 and H1 contains one 2d-convolutional layer only, with activation functions as sigmoid and linear respectively followed by batch-normalization as well. The output of the attention module is an embedded feature vector and then goes through a final fully-connected layer with softmax activation to become a 50-dimension vector indicating class-wise probabilities. 

\subsection{The Attention Block}
There are two branches in the attention block and the details are described below. In the attention branch, a 1-by-1 convolutional layer %with {\color{red} $1\times1$ filter size} % what is a 1x1 filter--is that the same as a scalar?
and sigmoid activation is firstly applied to the output of the embedding block $\mathbf{X}^\prime\in\mathbb{R}^{C^\prime\times{T^\prime}\times{1}}$:
\begin{equation}\label{sigmoid}
    \mathbf{X}^{\prime}_S=Sigmoid(Conv^{1\times1}(\mathbf{X}^{\prime}))
\end{equation} 
where $\mathbf{X}^{\prime}_S$ has the same dimensions as $\mathbf{X}^{\prime}$ and $Conv$ stands for a convolutional layer. Then, the attention vectors $\mathbf{A}\in\mathbb{R}^{C^\prime\times{T^\prime}\times{1}}$ are obtained by performing normalization along the $T^\prime$ dimension to make sure the weights sum up to 1, therefore acting like probabilities:
\begin{equation}\label{norm}
    \mathbf{A}(c,t,1)=\frac{\mathbf{X}^{\prime}_S(c,t,1)}{\sum_{t=1}^{T^{\prime}}\mathbf{X}^{\prime}_S(c,t,1)}
\end{equation}
where $c\in[1,C^\prime]$ and $t\in[1,T^\prime]$ are indices of the channel and time dimension. This step guaranteed the attention is applied temporally and different across channels. In the other branch, $\mathbf{X}^\prime$ goes through the same convolutional layer but without activation to become $\mathbf{X}_{L}^\prime$ with the same size: % can you explain why?
\begin{equation}\label{linear1}
    \mathbf{X}^{\prime}_L=Conv^{1\times1}(\mathbf{X}^{\prime})
\end{equation} 
Afterwards, the attention weighted feature vectors
$\mathbf{X}_{A}^\prime\in\mathbb{R}^{C^\prime\times{T^\prime}\times{1}}$ are obtained by element-wise multiplication between $\mathbf{A}$ and $\mathbf{X}_{L}^\prime$:
\begin{equation}\label{linear2}
    \mathbf{X}^{\prime}_A=\mathbf{X}^{\prime}_L \circ \mathbf{A}
\end{equation}
where $\circ$ stands for Hadamard product. In the end, a single hidden vector $\mathbf{H}\in\mathbb{R}^{C^\prime\times{1}}$ is obtained by summing up $\mathbf{X}_{A}^\prime$ along the time dimension and squeeze the frequency dimension before it is fed into a final fully connected layer for final classification:
\begin{equation}\label{linear3}
    \mathbf{H} = Squeeze_{f}(\sum_{t=1}^{T^\prime}\mathbf{X}_{A}^{\prime}(:,t,1))
\end{equation} 
where $t\in[1,T^\prime]$ is again the time index and $Squeeze_f()$ stands for the dimension squeeze operation that removes the frequency dimension. Batch normalization and ReLU activation are applied at the end of the attention block. In addition, a dropout with rate 0.3 is applied before entering the final fully connected layer.

\section{Experimental Setup}
\label{sec:expsetup}
\subsection{Datasets}
\subsubsection{ESC-50 and ESC-10}
ESC-50 \cite{piczak2015esc} has 2000 5-second-long audio clips with a sample rate of 44.1kHz, which are organized into 50 balanced classes. ESC-10 serves as a small proof-of-concept subset of 10 classes selected from the main dataset \cite{piczak2015esc}. Both datasets are prearranged into 5 folds and the final classification performance is measured by taking the average of all 5-fold cross-validation accuracies.

To reduce overfitting, data augmentation was performed on each training sample by applying random time shifting, pitch shifting and adding Gaussian noise, resulting in 3 variants for each sample. For random time shifting, the maximum duration to be shifted is 2.5 seconds, which is half the length of an audio sample. The time shifts are only delays so that the information of some samples that only contain transient sound events at the beginning will not be eliminated. For pitch changing, the function $pitch\_shift$ of $librosa$ was used and the pitch factor was set to be randomly chosen between -4 and 4. Finally, Gaussian noise values sampled from a standard normal distribution were added to clean audio samples after multiplying a noise factor of 0.01. After augmentation, the number of samples in total is increased to 8000.

\subsubsection{DCASE 2018 task1A and 2019 task1A}
The development sets from the DCASE 2018 task1A \cite{mesaros2018multi} and the DCASE 2019 task1A \cite{mesaros2018multi} are used. Both of them contain 10-second-long audio clips with a 48kHz sampling rate from 10 classes. The DCASE 2018 task1A development set has 6122 clips for training and 2518 clips for validation, and the DCASE 2019 task1A development set has 9185 clips for training and 4185 for validation. No data augmentation was performed on these two datasets.

\subsection{Data Preparation}
The log-mel spectrograms of all audio samples from the above datasets were extracted with their original sample rates kept. In addition, the deltas and delta-deltas of these log-mel spectrograms were also computed to obtain time-dependent information. The frame length and hop size for short-time-Fourier-transform are 1024 and 512, and the number of mel-bands is 128. All the features were extracted using the $librosa$ implementation. Instead of concatenating them together, these three feature maps were appended as three different channels to be the convolutional neural network inputs.

\subsection{Single-channel Attention and Non-attention Models}
\begin{figure}
\centering
\includegraphics[scale=0.11]{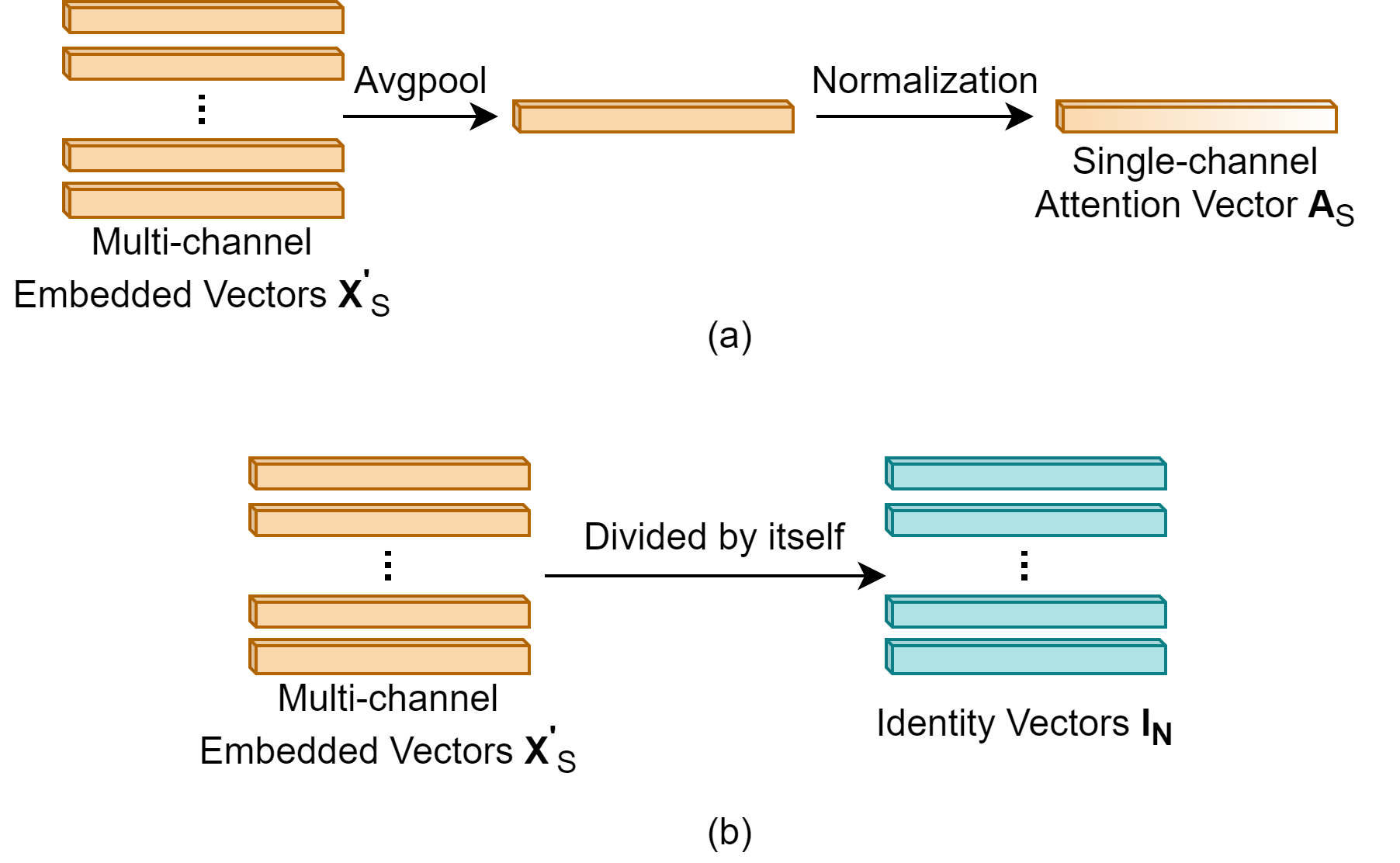}
\caption{\small (a) The architecture of single-channel temporal attention. (b) Non-attention architecture by setting all attention weights to 1.} 
\vspace{-0.1in}
\label{fig:single_and_non}
\end{figure}

To demonstrate the effectiveness of the proposed multi-channel temporal attention model, we also tested the performance of both the single-channel temporal attention model and the non-attention model using the same embedding block. \textbf{Fig. \ref{fig:single_and_non}} shows the architectures of the upper branch in the attention block corresponding to these two models, respectively. In the single-channel case, an average pool is applied on the channel dimension of $\mathbf{X}^{\prime}_S$ before the normalization. The final attention vector $\mathbf{A}_S\in\mathbb{R}^{{T^\prime}\times{1}}$ will be element-wisely multiplied with each $T^\prime\times{1}$ feature vector in $\mathbf{X}_{L}^\prime\in\mathbb{R}^{C^\prime\times{T^\prime}\times{1}}$. In the non-attention case, $\mathbf{X}^{\prime}_S$ is simply divided by itself element-wisely to form a bunch of identity vectors and the effect of attention will be removed in this way. These two models also have the same number of parameters as the multi-channel model so that these three models are comparable.

\subsection{Training the Network}
We trained the network using the Adam optimizer with cross-entropy loss. The learning rate was chosen to begin with 0.001 and decayed by a factor of 0.5 if the training losses of two consecutive epochs did not decrease. The mini-batch size is 50 with the number of epochs as 50 for each of the 5-fold cross validation. The number of hidden channels is 512 and the dropout rate is 0.3. The network was implemented with PyTorch and trained on an NVIDIA RTX2080Ti.

\section{Results and Discussion}
\label{sec:expresults}

To illustrate the effect of multi-channel temporal attention, experiments were performed using MCTA, a single-channel attention model, and a non-attention model. Each model was trained 5 times and the average classification accuracies and standard deviations were reported. When compared with other work, some results were reproduced by us by adapting the code available from the authors in our own experimental settings and the results were marked by *. \textbf{Table \ref{table:cmpesc}} shows the performance of our proposed model and other state-of-the-art methods including some recent attention models on the ESC datasets (ESC-50 and ESC-10). It was observed that temporal attention did improve the classification results by a noticeable margin and multi-channel temporal attention performed better than single-channel attention only. Our MCTA-CNN model outperformed the CNN baseline \cite{piczak2015environmental} by a very large margin, and outperformed the pretrained model on Audioset \cite{gemmeke2017audio} by Kumar \cite{kumar2018knowledge}. The FBE-ConvRBM \cite{sailor2017unsupervised} model uses filter-bank learning to extract features and applies system fusion to obtain the final results, and our model outperforms it with just a single model.  More importantly, our model performs better than multi-stream temporal attention \cite{li2019multi}, ACRNN \cite{zhang2019attention}, channel-temporal attention network \cite{zhang2019attention}, and TS-CNN10 \cite{wang2020environmental}, which all represent temporal attention using only one single vector.

\begin{table}
\centering
\footnotesize
\begin{tabular}{|c|c|c|}
\hline
Method/feature sets & ESC-50  & ESC-10  \\ \hline
Piczak CNN \cite{piczak2015environmental} & 64.50\% & 80.50\%\\
CNN pretrained on Audioset \cite{kumar2018knowledge} & 83.50\% & -\\
FBEs$\oplus$ConvRBM-BANK \cite{sailor2017unsupervised} & 86.50\% & -\\
Multi-stream temporal attention \cite{li2019multi} & 84.00\% & 94.20\%\\
ACRNN \cite{zhang2019attention} & 86.10\% & 93.70\%\\
Channel-temporal attention \cite{zhang2019learning} & 85.80\% & 94.00\%\\
TS-CNN10 \cite{wang2020environmental}* & 85.78$\pm$0.40\% & 94.25$\pm$0.28\%\\ \hline
\textbf{Non-attention(ours)} & 81.74$\pm$0.56\% & 90.95$\pm$0.40\% \\
\textbf{Single-channel(ours)} & 85.24$\pm$0.12\% & 93.15$\pm$0.20\% \\
\textbf{MCTA-CNN (ours)} &  \textbf{87.05$\pm$0.18}\% &  \textbf{94.45$\pm$0.24}\%\\ \hline
\end{tabular}
\caption{\small Comparison of classification accuracy of ESC-50 and ESC-10 datasets with other work. The $\oplus$ sign means system combination. The * denotes that the results were reproduced by us. The original results reported in \cite{wang2020environmental} were 88.6\% and 95.8\% for ESC-50 and ESC-10 respectively.}
\label{table:cmpesc}
\end{table}

Our model was also tested on the two DCASE acoustic scene classification (ASC) datasets described above, and the results are shown in \textbf{Table \ref{table:cmpdcase}}. It is shown that MCTA can also improve the performance of ASC as well. Our model outperformed several popular attention-based CNN models including the self-attention model \cite{wang2018self} and the Atrous-CNN model \cite{ren2019attention}, which were both initially designed for acoustic scene classification. Additionally, the number of parameters of our model is 1.47M, which is much smaller than TS-CNN10 \cite{wang2020environmental} (4.98M) and Atrous-CNN \cite{ren2019attention} (4.36M).

\begin{table}
\centering
\small
\begin{tabular}{|c|c|c|}
\hline
Methods & DCASE2018 1A  & DCASE2019 1A  \\ \hline
Self-attention \cite{wang2018self} & 70.81\% & -\\
Atrous-CNN \cite{ren2019attention}* & 69.07$\pm$0.59\% & 69.24$\pm$0.99\%\\
TS-CNN10 \cite{wang2020environmental}* & 69.68$\pm$0.60\% & 69.59$\pm$0.68\%\\
 \hline
\textbf{Non-attention(ours)} & 70.98$\pm$0.66\% & 74.60$\pm$0.32\% \\
\textbf{Single-channel(ours)} & 71.92$\pm$0.75\% & 74.88$\pm$0.40\% \\
\textbf{MCTA-CNN (ours)} &  \textbf{72.40$\pm$0.38}\% &  \textbf{75.71$\pm$0.28}\%\\ \hline
\end{tabular}
\caption{\small Comparison of classification accuracy of DCASE 2018 task1A and DCASE 2019 task1A datasets with other work. The * denotes that the results were reproduced by us. In \cite{wang2020environmental} the original results were 68.7\% and 70.6\% for DCASE 2018 and 2019 respectively. In \cite{ren2019attention}, the original result for DCASE 2018 was 72.7\% and no DCASE 2019 result was given.}
\label{table:cmpdcase}
\end{table}

\begin{figure}[ht]
\centering
\includegraphics[width=\linewidth]{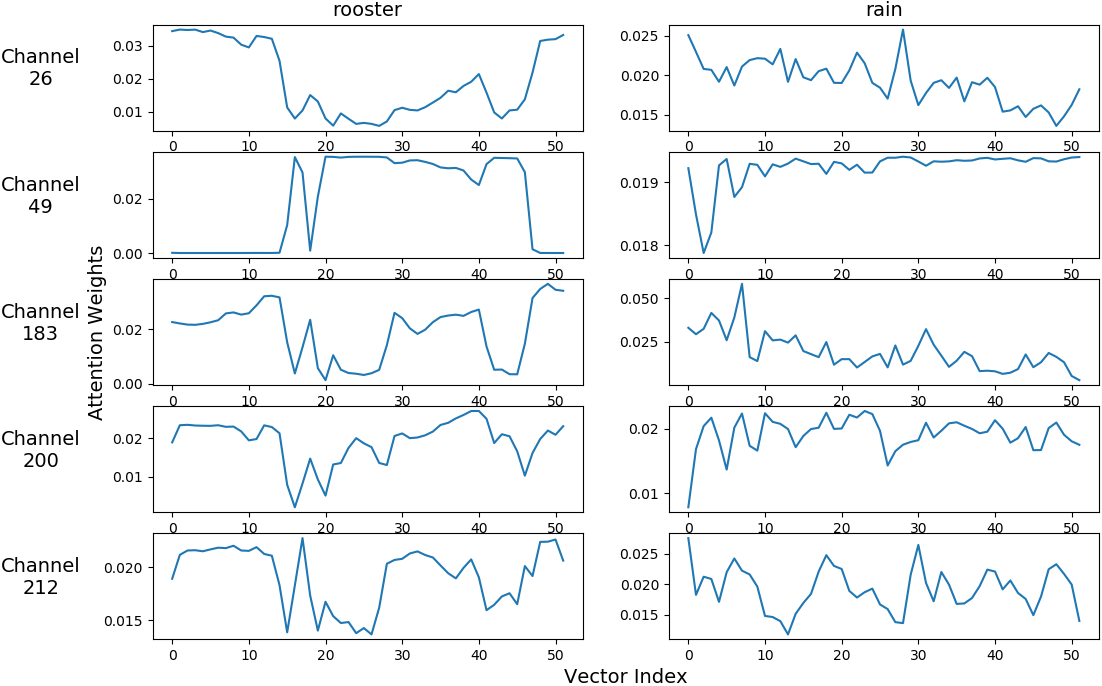}
\caption{\small 5 random attention vectors for rooster and rain.} 
\label{fig:att_vec}
\end{figure}

\textbf{Fig. \ref{fig:att_vec}} shows the attention vectors from 5 random channels (out of the 512 total in $\mathbf{A}\in\mathbb{R}^{512\times{52}\times{1}}$) for two classes in ESC-50, namely rooster and rain. It can be seen that within each class, the attention vectors from different channels can be very different, both in shape and magnitude, which further supports our proposition that a unique attention vector for each channel is beneficial.

\section{Conclusions and Future Work}
\label{sec:conclusions}
In this work, we propose a multi-channel temporal attention (MCTA) model for environmental sound classification. Our model can fully exploit the temporal information contained in all channels and is relatively lighter. The experimental results on ESC and ASC datasets validate the performance of MCTA. In the future, we will further test our model on larger datasets such as Audioset.

% Below is an example of how to insert images. Delete the ``\vspace'' line,
% uncomment the preceding line ``\centerline...'' and replace ``imageX.ps''
% with a suitable PostScript file name.
% -------------------------------------------------------------------------

%\vfill
%\pagebreak

\vfill\pagebreak

% \bibliographystyle{IEEEbib}
% {\small
% \bibliography{strings,refs}
% }

\end{document}